\documentclass[journal]{IEEEtran}
\makeatletter\if@twocolumn\PassOptionsToPackage{switch}{lineno}\else\fi\makeatother

      \makeatletter
\usepackage{wrapfig}
\newcounter{aubio}

\long\def\bioItem{%
\@ifnextchar[{\@bioItem}{\@@bioItem}}

\long\def\@bioItem[#1]#2#3{
 \stepcounter{aubio}
 \expandafter\gdef\csname authorImage\theaubio\endcsname{#1}
 \expandafter\gdef\csname authorName\theaubio\endcsname{#2}
 \expandafter\gdef\csname authorDetails\theaubio\endcsname{#3}
}

\long\def\@@bioItem#1#2{
 \stepcounter{aubio}
 \expandafter\gdef\csname authorName\theaubio\endcsname{#1}
 \expandafter\gdef\csname authorDetails\theaubio\endcsname{#2}
}

\newcommand{\checkheight}[1]{%
  \par \penalty-100\begingroup%
  \setbox8=\hbox{#1}%
  \setlength{\dimen@}{\ht8}%
  \dimen@ii\pagegoal \advance\dimen@ii-\pagetotal
  \ifdim \dimen@>\dimen@ii
    \break
  \fi\endgroup}

\def\printBio{%
  \@tempcnta=0
   \loop
     \advance \@tempcnta by 1
     \def\aubioCnt{\the\@tempcnta}
     \setlength{\intextsep}{0pt}%
     \setlength{\columnsep}{10pt}%
     \expandafter\ifx\csname authorImage\aubioCnt\endcsname\relax%
      \else%
       \checkheight{\includegraphics[height=1.25in,width=1in,keepaspectratio]{\csname authorImage\aubioCnt\endcsname}}
        \begin{wrapfigure}{l}{25mm}
         \includegraphics[height=1.25in,width=1in,keepaspectratio]{\csname authorImage\aubioCnt\endcsname}
        \end{wrapfigure}\par
      \fi
     \noindent\textbf{\csname authorName\aubioCnt\endcsname}\csname authorDetails\aubioCnt\endcsname \par\bigskip
      \ifnum\@tempcnta < \theaubio
   \repeat
   }
\makeatother

\usepackage[T1]{fontenc}
\usepackage{amsmath}
\usepackage{times}
\usepackage{graphicx}
\usepackage{multirow}
\usepackage[none]{hyphenat}
\usepackage{float}
\usepackage{subfig}
\ifCLASSINFOpdf
\else
\fi
\usepackage{amsmath}
\usepackage[cmintegrals]{newtxmath}
\usepackage{url,multirow,morefloats,floatflt,cancel,tfrupee}
\makeatletter

\AtBeginDocument{\@ifpackageloaded{textcomp}{}{\usepackage{textcomp}}}
\makeatother
\usepackage{colortbl}
\usepackage{xcolor}
\usepackage{pifont}
\usepackage[nointegrals]{wasysym}
\makeatletter

\def\mcWidth#1{\csname TY@F#1\endcsname+\tabcolsep}

\def\cAlignHack{\rightskip\@flushglue\leftskip\@flushglue\parindent\z@\parfillskip\z@skip}
\def\rAlignHack{\rightskip\z@skip\leftskip\@flushglue \parindent\z@\parfillskip\z@skip}

\usepackage{ifxetex}
\ifxetex\else\if@twocolumn\usepackage{dblfloatfix}\fi\fi

\AtBeginDocument{
\expandafter\ifx\csname eqalign\endcsname\relax
\def\eqalign#1{\null\vcenter{\def\\{\cr}\openup\jot\m@th
  \ialign{\strut$\displaystyle{##}$\hfil&$\displaystyle{{}##}$\hfil
      \crcr#1\crcr}}\,}
\fi
}

\AtBeginDocument{%
  \@ifpackageloaded{endfloat}%
   {\renewcommand\efloat@iwrite[1]{\immediate\expandafter\protected@write\csname efloat@post#1\endcsname{}}}{\newif\ifefloat@tables}%
}%

\def\BreakURLText#1{\@tfor\brk@tempa:=#1\do{\brk@tempa\hskip0pt}}
\let\lt=<
\let\gt=>
\def\processVert{\ifmmode|\else\textbar\fi}

\@ifundefined{subparagraph}{
\def\subparagraph{\@startsection{paragraph}{5}{2\parindent}{0ex plus 0.1ex minus 0.1ex}%
{0ex}{\normalfont\small\itshape}}%
}{}

\newcommand\role[1]{\unskip}
\newcommand\aucollab[1]{\unskip}
  
\@ifundefined{tsGraphicsScaleX}{\gdef\tsGraphicsScaleX{1}}{}
\@ifundefined{tsGraphicsScaleY}{\gdef\tsGraphicsScaleY{.9}}{}
\def\checkGraphicsWidth{\ifdim\Gin@nat@width>\linewidth
	\tsGraphicsScaleX\linewidth\else\Gin@nat@width\fi}

\def\checkGraphicsHeight{\ifdim\Gin@nat@height>.9\textheight
	\tsGraphicsScaleY\textheight\else\Gin@nat@height\fi}

\def\fixFloatSize#1{}
\let\ts@includegraphics\includegraphics

\def\inlinegraphic[#1]#2{{\edef\@tempa{#1}\edef\baseline@shift{\ifx\@tempa\@empty0\else#1\fi}\edef\tempZ{\the\numexpr(\numexpr(\baseline@shift*\f@size/100))}\protect\raisebox{\tempZ pt}{\ts@includegraphics{#2}}}}

\AtBeginDocument{\def\includegraphics{\@ifnextchar[{\ts@includegraphics}{\ts@includegraphics[width=\checkGraphicsWidth,height=\checkGraphicsHeight,keepaspectratio]}}}

\DeclareMathAlphabet{\mathpzc}{OT1}{pzc}{m}{it}

\def\URL#1#2{\@ifundefined{href}{#2}{\href{#1}{#2}}}

\def\UrlOrds{\do\*\do\-\do\~\do\'\do\"\do\-}%
\g@addto@macro{\UrlBreaks}{\UrlOrds}

\@ifundefined{quoteAttrib}
	{}
	{}

\@ifundefined{titlequoteAttrib}
	{}{}

\newenvironment{title-quote}
	{\list{}{\fontsize{10pt}{12pt}\selectfont\leftmargin.5in\itshape\rightmargin\leftmargin}%
  \item\relax}
  {\endlist}

\makeatother

\makeatletter
\AtBeginDocument{\@ifpackageloaded{longtable}{%
\def\LT@makecaption#1#2#3{%
  \LT@mcol\LT@cols c{\hbox to\z@{\hss\parbox[t]\LTcapwidth{%
    \sbox\@tempboxa{#1{#2: } #3}%
    \ifdim\wd\@tempboxa>\hsize
      #1{#2: }\textsc{#3}%
    \else
      \hbox to\hsize{\hfil\box\@tempboxa\hfil}%
    \fi
    \endgraf\vskip\baselineskip}%
  \hss}}}
}{}}
\makeatother
\begin{document}
\raggedbottom
%
%
%
\title{A High Sensitivity Fourier Transform Spectrometer for Cosmic Microwave Background Observations}
%
%
\author{Javier De Miguel-Hern\'andez, Roger J. Hoyland, Mar\'ia F. G\'omez Re\~nasco, \newline J. Alberto Rubi\~no-Mart\'in, Teodora A. Viera-Curbelo}
\maketitle
%
%
\begin{abstract}
 The QUIJOTE Experiment was developed to study the polarization in the Cosmic Microwave Background (CMB) over the frequency range of 10-50 GHz. Its first instrument, the Multi Frequency Instrument (MFI), measures in the range 10-20 GHz which coincides with one of the naturally transparent windows in the atmosphere. The Tenerife Microwave Spectrometer (TMS) has been designed to investigate the spectrum between 10-20 GHz in more detail. The MFI bands are 2 GHz wide whereas the TMS bands will be 250 MHz wide covering the complete 10-20 GHz range with one receiver chain and Fourier spectral filter bank. It is expected that the relative calibration between frequency bands will be better known than the MFI channels and that the higher resolution will provide essential information on narrow band interference and features such as ozone. The TMS will study the atmospheric spectra as well as provide key information on the viability of ground-based absolute spectral measurements. Here the novel Fourier transform spectrometer design is described showing its suitability to wide band measurement and $\sqrt{N}$ advantage over the usual scanning techniques.
\end{abstract}
%
%

\section{Introduction}

There are several microwave instruments that have been designed for CMB spectral measurement (e.g. \cite{1994ApJ...420..457F}, \cite{2007A&A...464..795H}, \cite{2011JCAP...07..025K}, \cite{2011ApJ...730..138S}). The present plan is to extend the goals of the QUIJOTE-CMB Experiment (\cite{2015MNRAS.452.4169G}, \cite{2017MNRAS.464.4107G}, \cite{2010ASSP...14..127R}, \cite{2012SPIE.8444E..2YR}) by building a new microwave spectrometer in the frequency range 10-20 GHz. The measurement of the CMB spectrum to the accuracy that will reveal interesting science is particularly difficult because there are many uncertainties that lead to effects of the order of this level. The instrumental thermal-noise due to finite integration times or shot-noise, intrinsic to the CMB photons distribution can be mitigated by daily stacking of a given sky region data since they are randomn effects. Systematics given by pointing calibration,  beam characterization, gain and bandwidth mismatch, component non-ideality, etc., become the largest non negligible contribution to uncertainty as the randomn noise is reduced by integration (e.g., \cite{LFI}, \cite{2017ITAP...65..644K}, \cite{2012RaSc...47.0K06R}, \cite{Winkel}). The systematics generally lie below the noise level through daily observation and can only be seen after various days of stacking data.

\begin{figure*}[t]
\centering
\includegraphics[width=150mm]{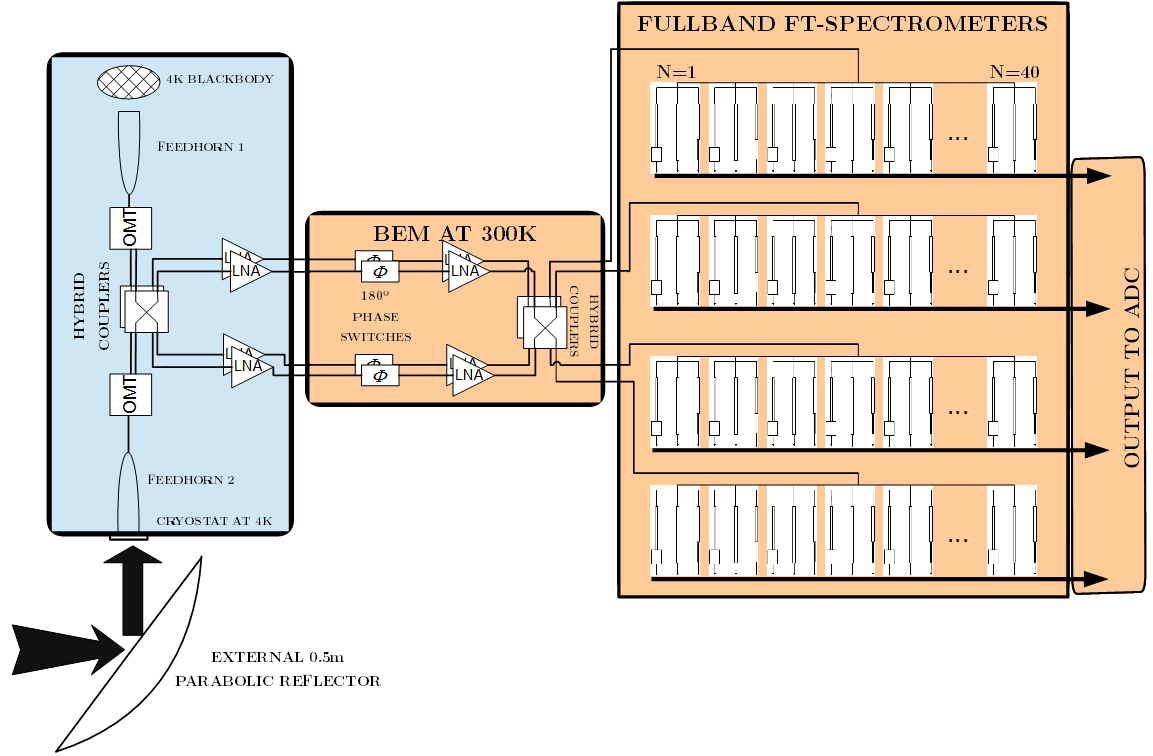}
\caption{A schematic diagram of the microwave spectrometer.}
\label{fig_1}
\vspace{-\baselineskip}
\end{figure*}

Fig. \ref{fig_1} shows the schematic diagram of a microwave spectrometer for measuring CMB spectra between 10 and 20 GHz. In order to achieve sufficient sensitivity the front end of a radiometer is cooled in a cryostat to 4-10 K. The cryogenic reference load, opto-mechanics and ultra-LNAs are cooled by a 4 K closed cycle helium gas cooler and maintained at a constant temperature slightly higher than 4 K by actively heating and controlling the coldstage.

The radiometer consists of a pair of novel ultra-wide band meta-horns \cite{Miguel_Hern_ndez_2019}, one pointed at a cold black body and the other looking out of the cryostat through a transparent window. These feedhorns are both followed by broadband 10-20 GHz waveguide orthomode transducers (OMTs) to couple the two linear polarizations. The polar outputs of both sky and load horns are fed to either arm of a pseudo-correlation radiometer which employs both cold and warm gain and $180^{\circ}$ broadband phase switch to switch the outputs. The correlating elements are $90^{\circ}$  hybrid couplers or ridged waveguide hybrid tees. Each correlator contains two similar low noise amplifiers (LNAs) with a noise temperature of less than 10 K followed by further amplification and a $180^{\circ}$ phase switch in each branch. The phase switches are housed in the back end module (BEM). Finally the load and sky signals are decorrelated in a similar $90^{\circ}$ hybrid coupler or ridged waveguide hybrid tee to that of the input. The outputs of these two hybrids are sent to four identical novel spectrometer banks. The DC output is amplified and then digitized for post processing in an ADC unit. 

This spectrometer is to be mounted in a simple telescope mount capable of maintaining the cryostat at a given declination whilst scanning through the azimuth throughout the observation. In order to achieve the goals of the project in both sensitivity, angular resolution and systematics, the spectral resolution shall be no greater than 250 MHz for each sub band, the system temperature shall be no greater than 22 K at any frequency over the band and the sensitivity of the instrument for a given band must be 2 $\operatorname{mKs^{1/2}}$  or better. The dynamic range of the receiver system shall be from root mean square (RMS) noise at 1 second (1.26 mK or 10 times less than the noise at the switch rate of 100 Hz) to 100 K plus the noise temperature of the receiver (maximum). This gives plenty of margin for any measurements that will be made. There is need of a dynamic range of 10000 or 16 bits minimum in the ADC to adequately sample the noise and measure bright calibration objects in the sky. 

In general the scientific requirements of the instrument are very strict in terms of sensitivity and stability. The sensitivity is dependent mainly on the cold frontend amplifiers while the stability is guaranteed by intelligent instrument design. The goals required for the spectrometer are very high spectral stability with a resolution of 250 MHz. The loss and sensitivity are not issues in this part of the instrument because of the high gain stages in the frontend. The overall stability is guaranteed with fast electronic phase switches and a 1 mK stable cold load reference. The spectral stability of the spectrometer is obtained through the design. Each frequency point of a Fourier Transform Spectrometer (FTS)  is obtained by the combination of all the detector outputs and any variation in their gain will be reflected in all the frequency points equally. The spectrometer is required to reproduce fourier components of the passband. Any deviations from this ideal can be calibrated using a network analyzer (VNA) at the input and sweeping out a spectrally flat response so that the output of each element can be measured. These outputs can be compared and calibrated against an FTS. No size and weight requirements are, in principle, relevant, since the mounting of the spectrometer has been originally designed for much larger and heavier instruments like the Quijote's MFI and Thirty Gigahertz Instrument (TGI).

In order to design an instrument able to measure the CMB spectral density function a new concept of FTS based on the technology of \textit{Instantaneous Frequency Measurement} (IFM) receivers has been designed. IFM were used in military technology with the objective of determining the frequency of a received signal, as well as in radar applications. The history and main features of these devices can be found by consulting the references of this document \cite{EAST}.

Thus, an elegant and simple three-channel design of a \textit{Reflection Mode Discriminator} receiver for IFM (\cite{RHODES}, \cite{IFM}, \cite{THORNTON}) has been converted into an innovative wideband autocorrelator for a novel Fourier Transform based spectrometer.

In the following sections we present this novel concept of FT-Spectrometer for an instantaneous ultra-broadband 10-20 GHz with a modifiable 250 MHz spectral resolution. Since the circuit measures Fourier components of the entire band simultaneously, it will present a signal to noise ratio improved by a $\sqrt{N}$ factor compared with the regular scanning technique used in other microwave spectrometers, which permits the high sensitivity required for its astrophysical purpose to be obtained. This benefit is known as Fellgett's advantage or multiplex advantage \cite{fellgett1951theory}, and makes the TMS theoretically faster than previous spectrometers designed for CMB-measurement (COBE-FIRAS, PIXIE, ARCADE2). On the other hand, a broad instantaneous bandwidth of 10-20 GHz allows an increase of the sensitivity by a factor $\sqrt{\Delta \nu}$ at the cost of loosing spectral resolution if this were necessary.
Since the entire band of the experiment is measured simultaneously, the TMS relative calibration between bands is expected to be more stable and presents lower uncertainty than the other experiment taking CMB measurements in the same frequency band in the north hemisphere, the Multi Frequency Instrument (MFI) \cite{2012SPIE.8452E..33H}, which presents four bands centered at 11, 13, 17 and 19 GHz with 2 GHz bandwidth, so the higher resolution of the TMS in comparison with the MFI will provide important information on narrow band interferences and features such as ozone.

Alternative designs of planar IFM devices have been published (e.g. \cite{7067876}, \cite{Badran}, \cite{8467497}, \cite{Rahimpour}, \cite{souza}). However, even in the cases where a 2:1 bandwidth factor is reached, the tendency for these prototypes is to use 4-channel (or more) typologies, while the design here described presents the advantage of being a 3-channel device, saving one channel for each Fourier component. On the other hand, these alternative designs generally do not produce orthogonal Fourier components of the passband shape, so the conversion of these IFM receivers into a FT-Spectrometer is not possible. Another limitation in these designs in comparison with the receiver here presented is the presence of band-pass filters for the discrimination of the signal, because filters are generally limited in bandwidth. Broadband digital spectrometers are very popular too. They provide a complete way to derive high-resolution spectra from a given bandwidth. They are limited in bandwidth (presently 2GHz) due to the speed of FPGAs but this will increase as technology advances. Digital spectrometers trade high resolution with complexity (power consumption, weight, CPU time). For measurements that do not require such high resolution a much simpler system can be designed that is lighter, with low power consumption needing little processing which may be priority in certain spectrometers, e.g., space based spectrometers, multi-channel spectrometers or high frequency spectrometers.

One disadvantage of our design in comparison with the alternative designs is the presence of a 3 channel power divider, which is not as balanced as the regular 2-channel-3[dB] power dividers present in these. We expect that a 2-channel version of our device can be developed and presented in future works, and we also expect that the bandwidth of the device can be also increased based on simulations and preliminary measures.

\section{System Fundamentals}
\subsection{Background}


The Michelson spectrograph is based on the Michelson-Morley experiment principle. The light of the source is divided into two beams. One is reflected in a fixed mirror and the other in a movable mirror. The beams interfere, and by making signal measurements in many discrete positions of the moving mirror, the spectrum of the source can be reconstructed.

\begin{figure}[H]
\centering
\includegraphics[width=90mm]{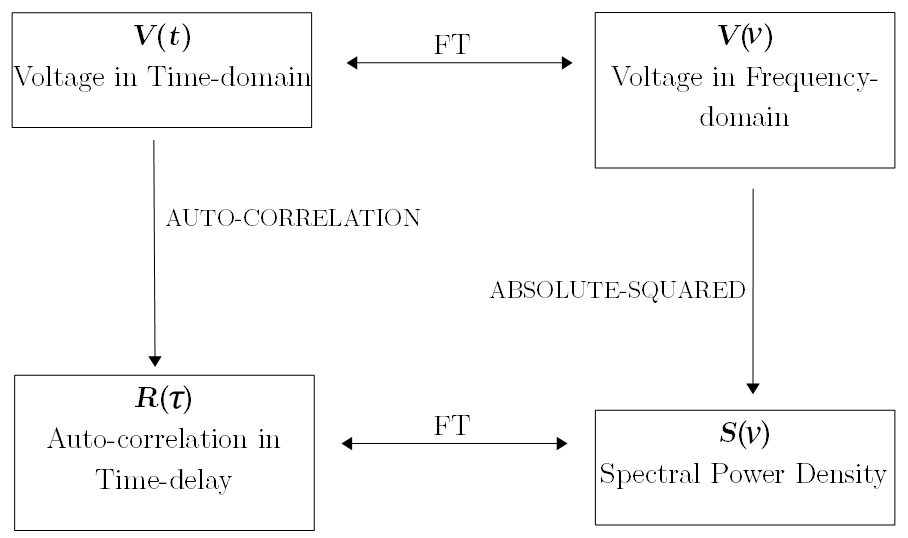}
\caption{Relation between the voltage and the autocorrelation function.}
\label{fig_2}
\end{figure}

The equivalent device of a Michelson spectrometer in radio astronomy is FTS. The RF autocorrelator supplants the movable mirror by an artificial delay obtained by dividing the received signal in N-1 parts that are forced to pass through lines of different electric length causing a path delay between them before being combined and measured. This is similar to an array of radio-telescopes applying interferometric techniques. The fundamentals for the signal reconstruction are shown in the Fig. \ref{fig_2} \cite{TOOLS}.

In Fig. \ref{fig_2} $R(\tau)$ is the autocorrelation function, $\nu$ is the frequency and $\tau$ is the \textit{time delay}-path so it is possible to write\newline

\begin{equation}
S(\nu) = \int_{-\infty}^{+\infty} R(\tau) \, e^{-2\pi i \tau \nu} d\tau \;
\label{equation_1}
\end{equation}

\begin{equation}
R(\tau) = \int_{-\infty}^{+\infty} S(\nu) \, e^{2\pi i \tau \nu} d\nu \;.
\label{equation_2}
\end{equation}

Since the autocorrelator samples the time average autocorrelation function over an integration time ($T$) with one or more delays and multipliers of the signal's voltage at two times $t$ and $t+\tau$ it is possible to write

\begin{equation}
R(\tau)=\frac{1}{2T} \int_{-T}^{+T}  v(t) \, v(t+\tau) \, dt=<\!v(t)  \; v(t+\tau)\!> \;.
\label{equation_3}
\end{equation}


\subsection{System Analysis}
\label{III}

\begin{figure*}[t]
\centering
\includegraphics[width=150mm]{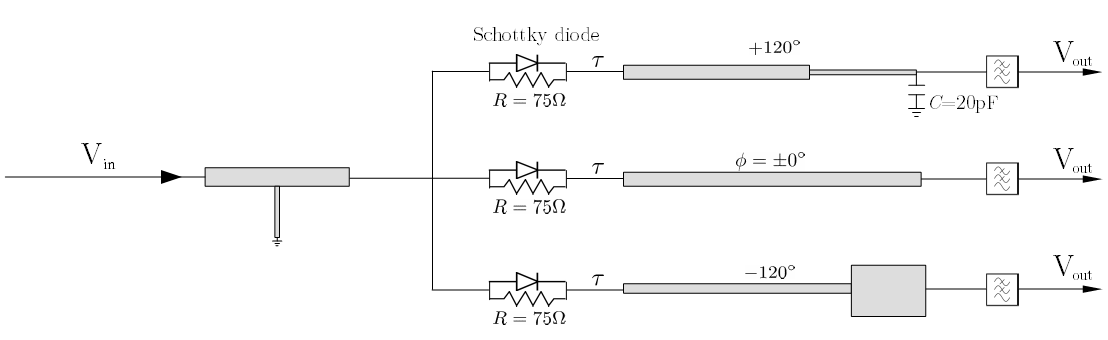}
\caption{N=1 spectral element schematic. The input signal on the left passes through a wide-band-tee \cite{KIM} in order to provide a DC ground before being divided into 3 circuits and passing through the $\tau$ delay-path  with differential $\phi$ phase-shifts generating three $V_{out}$ being filtered and sent to the acquisition system in order to be combined and processed.}
\label{fig_3}
\vspace{-\baselineskip}
\end{figure*}

The spectral element is schematically in presented in Fig. \ref{fig_3}. In this device, three channels with relative phase-shifts of 0, 120 and 240 degrees produce three signal-output voltages $V=1+\mathrm{cos}(\omega\tau+\phi)$, where $\omega=2\pi \nu$ and $\tau$ is the time delay.

Thus, each of the three channels introduces a phase shift in the signal $\omega \tau $ by using delay path length and relative phase shift between each channel of $\phi$ due to the transmission line impedance transformers. 
In this device, the 75 $\Omega$ resistance is mounted in parallel with a zero-biased diode detector. At frequencies where the electric length of a short circuited line is a $m$ multiple of $\lambda/2$ all the input power will be dissipated in the resistor, whereas at $n$ odd multiples of $\lambda/4$ the resistor and therefore the detector see no power because of a standing wave  effect. This situation is inverted for a line terminated with an open circuit.


\begin{figure}[H]
\centering
\includegraphics[width=70mm]{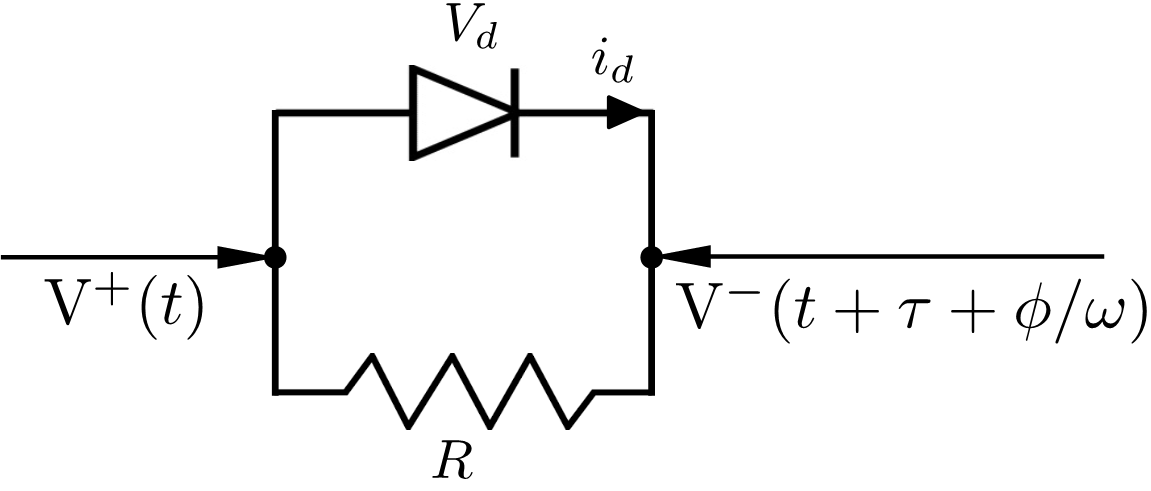}
\caption{Schematic of detection unit, where according to the quadratic response of the zero-biased schottky diode both the incident and the reflected waves are combined and detected, and so the autocorrelation function is measured.}
\label{fig_4}
\end{figure}

 Each schottky diode detector measures the power in the corresponding channel of the combined input and reflected waves  (after being reflected in the circuit of the device presented in Fig. \ref{fig_3}), allowing the autocorrelation function $R(\tau)$ given by (\ref{equation_3}) to be obtained. The detector output is subsequently processed through a data acquisition system. 
 The series equivalent resistance to the diode+resistor parallel circuit is approximately $R$ because the characteristic resistance of a Schottky diode is very high compared to the parallel resistor, this arrangement provides a quasi-perfect match to the circuit.
 Each element similar to the one shown in the Fig. \ref{fig_3} gives a point in the spectrum, so 40 elements (N=40) give a maximum resolution of 250 MHz spectrum in the band 10-20 GHz.

If the output from the detection units of each spectral element is obtained the full autocorrelation function in (\ref{equation_3}) can be derived. This is represented with the help of the Fig. \ref{fig_4} and it is analyzed in the following lines.  It is well known that the response $i_d$ per unit of area of the junction of a Schottky diode over its square law region for small signals is given by

\begin{equation}
\begin{split}
i_{d}[A/m^2]  =  I_s\left(e^{\frac{V_d}{\eta V_t}}-1\right) \simeq I_s\left[\frac{V_d}{\eta V_t} +\frac{1}{2} \left(\frac{V_d}{\eta V_t}\right)^2\right] ,
\end{split}
\label{equation_5}
\end{equation}

where a Taylor series has been applied and $V_d$ is the voltage across the diode, $\eta$ is a factor of order of the unity, $V_t$ is the thermal voltage\footnote{The thermal voltage is given by $V_t=K_BT/q\simeq 25\text{ mV}$ at $T\sim$ room temperature, where $K_B$ is the Boltzmann constant and $q$ is the elementary charge.} and $I_s$ is the \textit{reverse saturation current}, which depends on the physical characteristics of the diode (see, for example, \cite{1964pecm.book.....G} in order to know more). 

Since $V_d$ is the addition of the forward and the reflected waves, for the circuit in Fig. \ref{fig_4} we obtain

\begin{equation}
\begin{split}
V_d=V_-(t+\tau+\phi / \omega)- V_+(t)    =\\ 
 V_-\mathrm{cos}[\omega(t+\tau+\phi/\omega)] - V_+\mathrm{cos}(\omega t) \, .
\end{split}
\label{equation_6}
\end{equation}

After some algebra, the substitution of (\ref{equation_6}) into (\ref{equation_5}) yields

\begin{equation}
\begin{split}
i_{d}[A/m^2]  \simeq  
\overbrace{\frac{I_s(V_+^2+V_-^2)}{4(\eta V_t)^2}}^{i_{bias}\rightarrow DC_d} + \,\,\,\,\,\,\,\,\,\,\,\,\,\,\,\,\,\,\,\,\, \,\,\,\,\,\,\,\,\,\,\,\,\,\,\,\,   \\
\frac{I_s}{\eta V_t}\left\{V_-\mathrm{cos}[\omega(t+\tau+\phi/\omega)] - V_+\mathrm{cos}(\omega t) \right\} +\\
  \frac{I_s}{4(\eta V_t)^2}\left\{V_-^2\mathrm{cos}[2\omega(t+\tau+\phi/\omega)] + V_+^2\mathrm{cos}(2\omega t) \right\} -\\
  \frac{I_s}{(\eta V_t)^2}\underbrace{ \left\{  V_-\mathrm{cos}[\omega(t+\tau+\phi/\omega)]V_+\mathrm{cos}(\omega t) \right\}  }_{R_{\phi}(\tau)}   \, ,
\end{split}
\label{equation_7}
\end{equation}

where the overbraced term $DC_d$ is the bias current from the diode, and since for loss-less transmission lines $V^- = V^+$, the last term is the autocorrelation function $R_{\phi}(\tau)$ from (\ref{equation_3}) for each line of the 3-channel circuits represented in the Fig. \ref{fig_3}. Note that for any two elements of the N=40 element array bank, $\tau$ differs but the relative phase-shift delay $\phi$ has the same value for each respective one of the 3 channels.

For our circuit arrangement where all the high frequency terms are filtered before passing to the ADC all the time dependent terms vanishes, (\ref{equation_7}) can be simplified to

\begin{equation}
i_{d_{out}}[A/m^2]  \simeq \frac{I_s V^2}{2(\eta V_t)^2}\left[1-\mathrm{cos}(\omega\tau+\phi)\right] \;.
\label{equation_8}
\end{equation}

In (\ref{equation_8}) lossless transmission lines are assumed, so $V=V^+=V^-$.

\subsection{ Signal recovering and calibration}
\subsubsection{Theoretical Approach}
It is well known that any periodic function $f(t)$ can be approximated with a \textit{Fourier Complex Series}. That is

\begin{equation}
f( t) = \sum\limits_{n = 0 }^\infty  {{C_n}{e^{i \pi n\frac{t}{L}}}} \,,
\label{equation_20}
\end{equation}

where $n$ is the number of harmonic terms, $L$ is the amplitude of the interval or period and the $C_n$ terms are given by

\begin{equation}
{{C_n} = \frac{1}{{2L}}\int\limits_{ -L}^L {f\left( t \right){e^{-i \pi n\frac{t}{L}}}}dt } \,.
\label{equation_21}
\end{equation}

\begin{figure}[H]
\centering
\includegraphics[width=85mm]{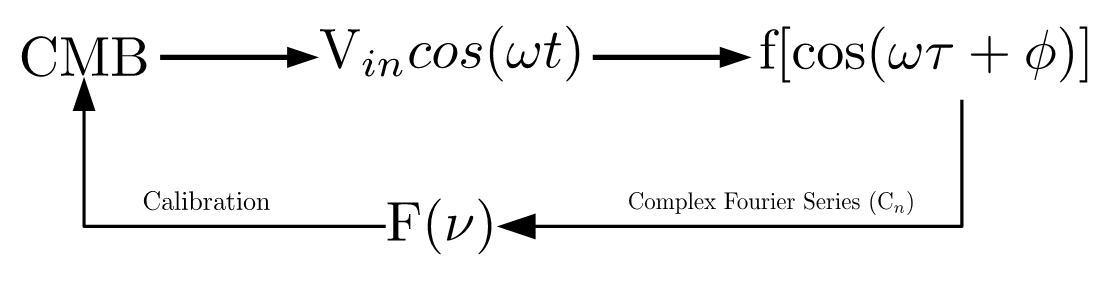}
\caption{Post-processing schematic.}
\label{fig_5}
\end{figure}

It can be seen that the expressions (\ref{equation_1}) and (\ref{equation_21}) are similar so the $C_n$ terms correspond to the discrete Fourier transform of the spectral density function\footnote{The approximation depends on the number of harmonics and their order.}. The relation is represented in Fig. \ref{fig_5}, where the incoming CMB produces a microwave electronic signal $V_{in} \mathrm{cos}(\omega t)$ once the photons are captured by the telescope-antenna system and the spectral element provides measurements of a $f[\mathrm{cos}(\omega \tau + \phi)]$ function in the time-delay domain which can be derived by the subtraction of pairs of output signals from each circuit\footnote{That is, the operations $i_{d_{out,a}}=i_{d_{out,1}}-i_{d_{out,2}}$,$i_{d_{out,b}}=i_{d_{out,1}}-i_{d_{out,3}}$ and $i_{d_{out,c}}=i_{d_{out,2}}-i_{d_{out,3}}$ outputs from (\ref{equation_8}).} in order to eliminate the DC offset, and this can be later converted to frequency domain and calibrated, recovering the spectral density function of the source, i.e., the CMB.

\subsubsection{Error and Calibration}
One of the challenges of the design stage is to keep the phase-shift between circuits flat over an ultra-wideband. Our iterative procedure of simulation-manufacturing-measurement reveals that the relative phase-shift error between any two circuits of the device could be limited to below 2 degrees \cite{Master}. This phase-shift error can increase in manufactured prototypes due to imperfections\footnote{In our iterative prototyping, where several procedures are hand-made, the relative phase-shift error is increased by a factor 2 or 3, but we think that it could be much lower by industrial techniques.}. Thus, it is interesting to estimate the total error introduced by differential phase-shift errors by introducing these error terms in (\ref{equation_8}) and normalizing for simplicity. This yields

\begin{equation}
\begin{cases} 

 V_{1_{\varepsilon}}=1-\mathrm{cos}(\omega \tau)\\

 V_{2_{\varepsilon}}=1-\mathrm{cos}(\omega \tau+2\pi/3+\phi_{\varepsilon_{12}}+\phi_{\varepsilon_{23}})\\

 V_{3_{\varepsilon}}=1-\mathrm{cos}(\omega \tau+4\pi/3-\phi_{\varepsilon_{23}}+\phi_{\varepsilon_{13}})\;,\\

\end{cases} \;
 \,\,\, \,\,\, 
\label{equation_22}
\end{equation}

where $V_1$ is considered the reference of the differential phase-shifts and $\phi_{\varepsilon_{ij}}$ is the relative phase-shift error between two $i$,$j$ circuits. From (\ref{equation_22}), and considering that $3DC=V1+V2+V3$ so $3DC_{\varepsilon}=V_{1_{\varepsilon}}+V_{2_{\varepsilon}}+V_{3_{\varepsilon}}$ it is possible to obtain, after some trigonometry, the analytical expression for the error in the estimation of the DC. This is

\begin{equation}
\begin{split}
 \varepsilon_{DC}=\frac{\mathrm{cos}(\omega \tau)}{2}[2-\mathrm{cos}(\phi_{\varepsilon_{12}}+\phi_{\varepsilon_{23}})-\mathrm{cos}(-\phi_{\varepsilon_{23}}+\phi_{\varepsilon_{13}})+
 \\\sqrt{3} \left(\mathrm{sin}(-\phi_{\varepsilon_{23}}+\phi_{\varepsilon_{13}})-\mathrm{sin}(\phi_{\varepsilon_{12}}+\phi_{\varepsilon_{23}}) \right)] +
 \\\frac{\mathrm{sin}(\omega \tau)}{2}\lbrace\sqrt{3}[\mathrm{cos}(\phi_{\varepsilon_{12}}+\phi_{\varepsilon_{23}})-\mathrm{cos}(-\phi_{\varepsilon_{23}}+\phi_{\varepsilon_{13}})]\\-\mathrm{sin}(\phi_{\varepsilon_{12}}+\phi_{\varepsilon_{23}})-\mathrm{sin}(-\phi_{\varepsilon_{23}}+\phi_{\varepsilon_{13}}) \rbrace
  \, .
\end{split}
\label{equation_23}
\end{equation}

\begin{figure}[H]
\centering
\includegraphics[width=90mm]{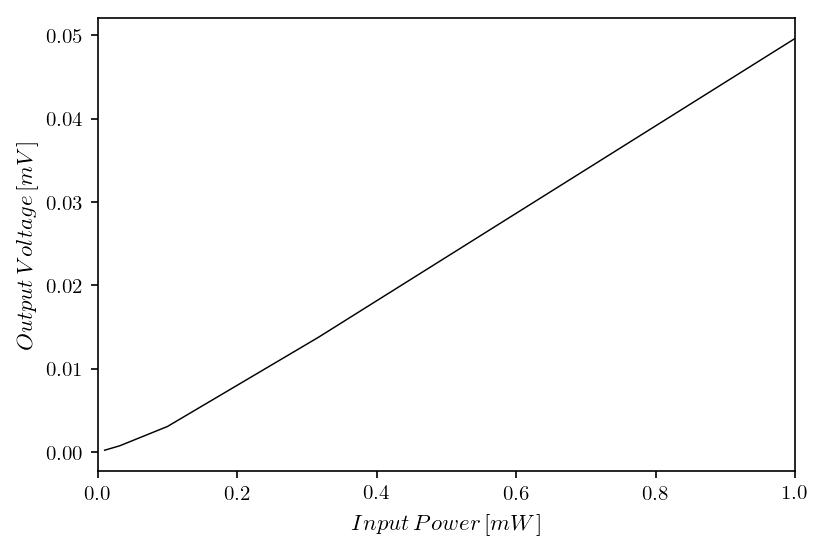}
\caption{Prototype linearity. Note that only small signals are measured in order to limit the input signal of the diode to values within the square law region.}
\label{fig_6}
\end{figure}

From (\ref{equation_23}), and assuming a lossless system, it can be established that the error in the estimation of the DC output of the channels is a function of the frequency ($\omega$), the delay-path ($\tau$) and the relative phase-shift error terms ($\phi_{\varepsilon_{ij}}$), which although lie within a narrow range are frequency dependent.
Simulations reveal that $\varepsilon_{DC}$ has values in the range 0-4\% for all the possible combinations of relative phase-shift errors $\phi_{\varepsilon_{ ij}}=\pm2^\circ$. Since the relative phase-shift error function is known and repeatable over the band, it can be calibrated\footnote{Together with other systematics.} because the system exhibits a high linearity and stability. The linearity is represented in the Fig. \ref{fig_6}. This allows us to experimentally calibrate the response by the same regular techniques used in the MFI instrument calibration, where a known input signal (laboratory or source in the sky) produce a unique and calibratable response. In the case of the MFI, calibration results are better than a 2\%, while the ongoing TGI calibration reveals that much lower uncertainties in observational calibrations are possible using ground based calibration sources.

\subsubsection{Uncertainty and Sensitivity}

 The theoretical uncertainty in terms of RMS of the minimal detectable temperature can be estimated by adapting the well known \textit{Dicke's ideal radiometer equation}. This is

\begin{equation}
\delta T=\frac{T_{sys}}{\sqrt{\Delta\nu_R \Delta t N }}\,.
\label{equation_24}
\end{equation}

The system temperature ($T_{sys}$) is a well established figure of merit in radioastronomy, $\Delta\nu_R$ is the bandwidth given by the spectral resolution and $N$ is the number of detectors integrating the entire instantaneous bandwidth at the same time for a $\sqrt{N}$ advantage. For our case where $T_{sys}\simeq $22 K, $\Delta\nu_R$=250 MHz and $N$=40 the expected sensitivity is 0.31 mKs$^{1/2}$ for each polarization.

\section{Experimental Setup}
\begin{figure}[H]
\centering
\includegraphics[width=90mm]{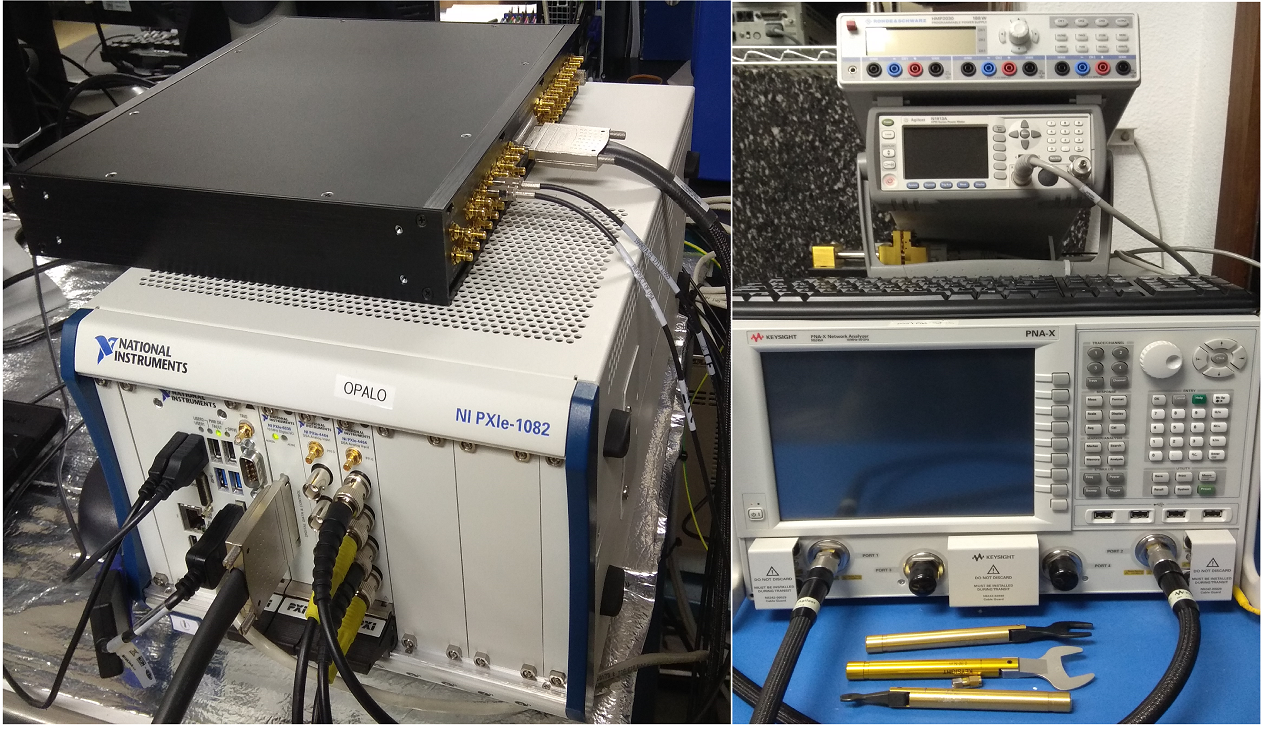}
\caption{Experimental test bench. DAS on the left, PNA-source on the right.}
\label{fig_7}
\end{figure}

\begin{figure}[H]
\centering
\includegraphics[width=70mm]{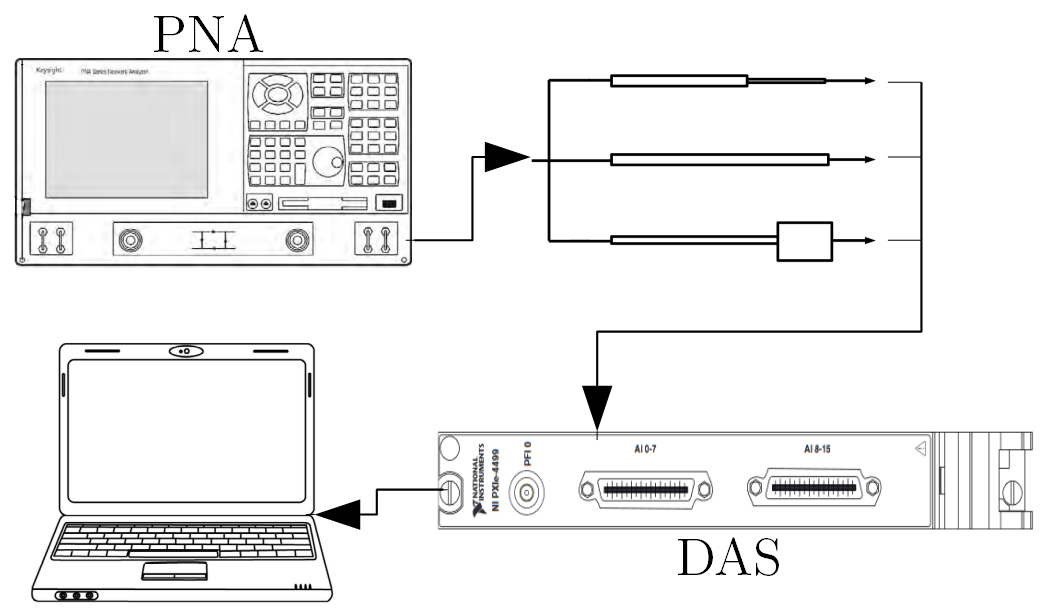}
\caption{Experimental setup scheme. The stabilized source of the PNA transmits the input signal to the circuit, whose output voltages are sent to a DAS, which stores them and sends them to a computer for the representation and post-processing of the data.}
\label{fig_0}
\end{figure}

A proof-of-concept detector unit prototype has been manufactured and measured in the laboratory. The experimental setup is described here. The Data Acquisition System (DAS) is shown in Fig. \ref{fig_7}, and consists of a PXIe-1082 chassis hosting a PXIe-8840 controller, a Dynamic Signal Acquisition analog input module PXIe-4465 with A/D converter of 24 bits to digitize the signal (ADC Delta-Sigma) and a 10 MHz Digital I/O device used to commutate the different phase switch angles (all modules from National Instruments). The DAS runs under Windows and a LabVIEW application has been developed which allows us to acquire the signal of the three channels at different sampling rates as well as to command phase switch commutations (not needed in this experiment). The application also communicates through ethernet, using DCOM technology, with the N5245A PNA-X Microwave Network Analyzer, from Keysight Technologies, allowing us to configure it for different bandwidths, powers, frequencies, etc., to generate the appropriate signal. This software allows the execution of a specific test for some fixed PNA parameters at a phase switch position and also it facilitates the execution of automatic tests by varying some PNA parameters in nested loops as well as the phase switch commutations allowing long terms tests to study the behavior of the signal. A schematic of the configuration is shown in Fig. \ref{fig_0}.

\begin{figure}[H]
\centering
\includegraphics[width=88mm]{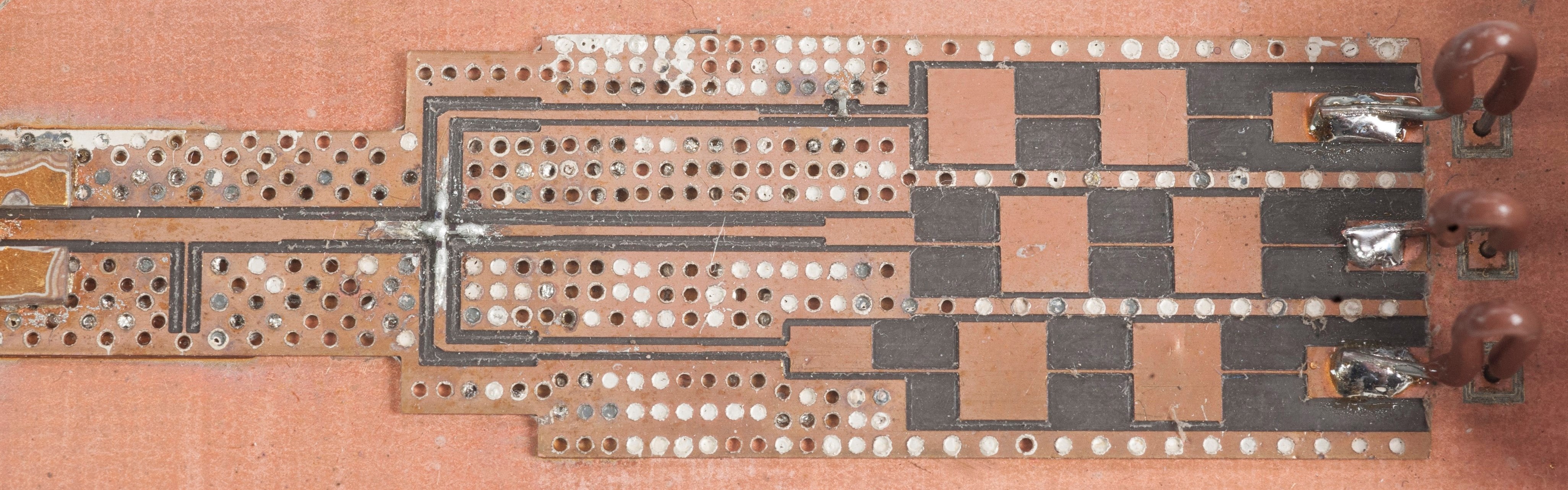}
\caption{Prototype of one detector unit. Approximate dimensions 18x60 mm fabricated in a ProtoMat-S100 machine in a Rogers-4003c laminate.}
\label{fig_8}
\end{figure}

\begin{figure}[H]
\centering
\includegraphics[width=92mm]{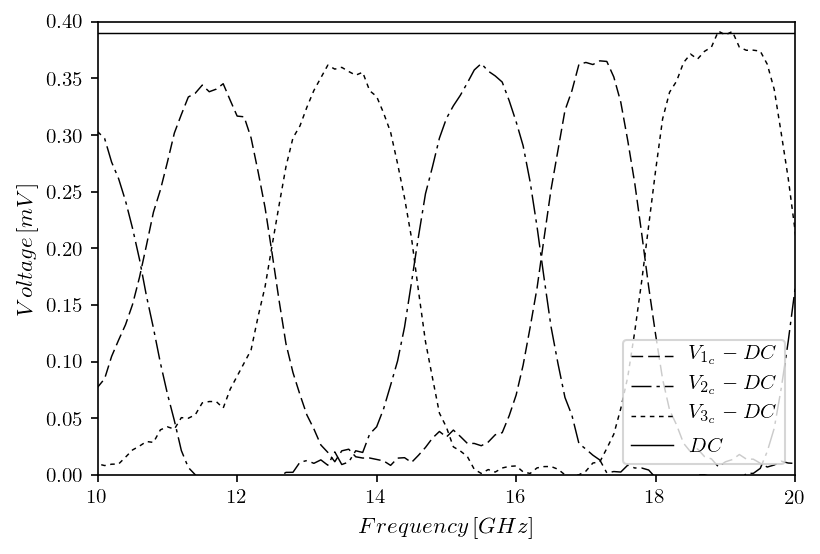}
\caption{Result of the measurements of the preliminary prototype.}
\label{fig_9}
\end{figure}

A picture of the prototype and the voltages from measurements are shown in figures \ref{fig_8} and \ref{fig_9} respectively. In Fig. \ref{fig_9} the measured corrected voltages are represented, \textit{i.e.}, since it is known that the perfect $DC$ voltage must present a plane horizontal voltage line, (\ref{equation_9}) can be applied

\begin{equation}
V_c=\frac{DC_{max}}{DC}V_m \;,
\label{equation_9}
\end{equation}

where $V_c$ is the corrected output voltage, $V_m$ is the measured voltage, $DC_{max}$ is the maximum DC voltage measured through the band and $DC$ is the DC voltage.

It is important to note that \textit{negative} values of voltage could not be measured because of the one-directional nature of a diode\footnote{This is common in detectors based on diode-type quadratic detectors.}. The square law detector measures power absorbed in the load, it is not phase sensitive to the incident or reflected wave but rather sensitive to the standing wave pattern formed by the reflected wave. The measurements in Fig. \ref{fig_9} are in agreement with this. This figure shows a recognizable standing wave pattern of the energy dissipated in each resistance of the three circuits, whose sum provides the $DC$ voltage necessary to obtain the spectral density function. A future experiment using the specific data acquisition system of the TMS would allow us to measure purer signals without spurious effects due to the limitations of the standard laboratory equipment. In the design phase, special attention must be paid to the forward signal divider, since it is also a virtual ground for the reflected voltages, and is sensitive to design or manufacturing errors. Improvements in the circuit given by trial and error measurements in the manufacture of dozens of detection units will allow us to perfectly balance the amplitudes of the 3 voltages coming from each circuit. On the other hand, industrial manufacturing of circuits could be much more repeatable and accurate than possible in a regular R\&D laboratory.

\section{Conclusions and Future Lines of Work}

The working principle of the novel FT-spectrometer has been demonstrated with very promising results.\newline The TMS will present a $\sqrt{N}$ integration-time advantage over previous microwave spectrometers, and it will have only static elements and will be small, strong and light, so it could be considered for future space missions. \newline Furthermore, this spectrometer is scalable to higher frequency bands maintaining the 2:1 bandwidth factor, so it could be used in future experiments exploring different spectral regions.

The experimental results were taken using a laboratory DAS. For experimental calibration tests a more advanced, sensitive and stable DAS will be necessary. This DAS is presently being designed  and it will allow us to fully complete the instrument.



\section*{Acknowledgment and Funding Information}

The authors wishes to acknowledge the support of all the technicians, engineers, scientists and administrative staff of the IAC and Quijote-CMB Experiment and to Miguel Briganty for the HD pictures. \newline


\bibliographystyle{IEEEtran}

\bibliography{IEEEabrv,IEEEexample}

\begin{thebibliography}{10}
\providecommand{\url}[1]{#1}
\csname url@samestyle\endcsname
\providecommand{\newblock}{\relax}
\providecommand{\bibinfo}[2]{#2}
\providecommand{\BIBentrySTDinterwordspacing}{\spaceskip=0pt\relax}
\providecommand{\BIBentryALTinterwordstretchfactor}{4}
\providecommand{\BIBentryALTinterwordspacing}{\spaceskip=\fontdimen2\font plus
\BIBentryALTinterwordstretchfactor\fontdimen3\font minus
  \fontdimen4\font\relax}
\providecommand{\BIBforeignlanguage}[2]{{%
\expandafter\ifx\csname l@#1\endcsname\relax
\typeout{** WARNING: IEEEtran.bst: No hyphenation pattern has been}%
\typeout{** loaded for the language `#1'. Using the pattern for}%
\typeout{** the default language instead.}%
\else
\language=\csname l@#1\endcsname
\fi
#2}}
\providecommand{\BIBdecl}{\relax}
\BIBdecl

\bibitem{1994ApJ...420..457F}
D.~J. {Fixsen}, E.~S. {Cheng}, D.~A. {Cottingham}, R.~E. {Eplee}, Jr.,
  T.~{Hewagama}, R.~B. {Isaacman}, K.~A. {Jensen}, J.~C. {Mather}, D.~L.
  {Massa}, S.~S. {Meyer}, P.~D. {Noerdlinger}, S.~M. {Read}, L.~P. {Rosen},
  R.~A. {Shafer}, A.~R. {Trenholme}, R.~{Weiss}, C.~L. {Bennett}, N.~W.
  {Boggess}, D.~T. {Wilkinson}, and E.~L. {Wright}, ``{Calibration of the COBE
  FIRAS instrument},'' \emph{ApJ}, vol. 420, pp. 457--473, Jan. 1994.

\bibitem{2007A&A...464..795H}
C.~M. {Holler}, T.~{Kaneko}, M.~E. {Jones}, K.~{Grainge}, and P.~{Scott}, ``{A
  6-12 GHz analogue lag-correlator for radio interferometry},'' \emph{A\&A},
  vol. 464, pp. 795--806, Mar. 2007.

\bibitem{2011JCAP...07..025K}
A.~Kogut \emph{et~al.}, ``{The Primordial Inflation Explorer (PIXIE): A Nulling
  Polarimeter for Cosmic Microwave Background Observations},'' \emph{JCAP},
  vol. 1107, p. 025, 2011.

\bibitem{2011ApJ...730..138S}
J.~{Singal}, D.~J. {Fixsen}, A.~{Kogut}, S.~{Levin}, M.~{Limon}, P.~{Lubin},
  P.~{Mirel}, M.~{Seiffert}, T.~{Villela}, E.~{Wollack}, and C.~A. {Wuensche},
  ``{The ARCADE 2 Instrument},'' \emph{ApJ}, vol. 730, p. 138, Apr. 2011.

\bibitem{2015MNRAS.452.4169G}
R.~{G{\'e}nova-Santos}, J.~A. {Rubi{\~n}o-Mart{\'{\i}}n}, R.~{Rebolo},
  A.~{Pel{\'a}ez-Santos}, C.~H. {L{\'o}pez-Caraballo}, S.~{Harper}, R.~A.
  {Watson}, M.~{Ashdown}, R.~B. {Barreiro}, B.~{Casaponsa}, C.~{Dickinson},
  J.~M. {Diego}, R.~{Fern{\'a}ndez-Cobos}, K.~J.~B. {Grainge}, C.~M.
  {Guti{\'e}rrez}, D.~{Herranz}, R.~{Hoyland}, A.~{Lasenby},
  M.~{L{\'o}pez-Caniego}, E.~{Mart{\'{\i}}nez-Gonz{\'a}lez}, M.~{McCulloch},
  S.~{Melhuish}, L.~{Piccirillo}, Y.~C. {Perrott}, F.~{Poidevin},
  N.~{Razavi-Ghods}, P.~F. {Scott}, D.~{Titterington}, D.~{Tramonte},
  P.~{Vielva}, and R.~{Vignaga}, ``{QUIJOTE scientific results - I.
  Measurements of the intensity and polarisation of the anomalous microwave
  emission in the Perseus molecular complex},'' \emph{Monthly Notices of the
  Royal Astronomical Society (MNRAS)}, vol. 452, pp. 4169--4182, Oct. 2015.

\bibitem{2017MNRAS.464.4107G}
R.~{G{\'e}nova-Santos}, J.~A. {Rubi{\~n}o-Mart{\'{\i}}n},
  A.~{Pel{\'a}ez-Santos}, F.~{Poidevin}, R.~{Rebolo}, R.~{Vignaga}, E.~{Artal},
  S.~{Harper}, R.~{Hoyland}, A.~{Lasenby}, E.~{Mart{\'{\i}}nez-Gonz{\'a}lez},
  L.~{Piccirillo}, D.~{Tramonte}, and R.~A. {Watson}, ``{QUIJOTE scientific
  results - II. Polarisation measurements of the microwave emission in the
  Galactic molecular complexes W43 and W47 and supernova remnant W44},''
  \emph{Monthly Notices of the Royal Astronomical Society (MNRAS)}, vol. 464,
  pp. 4107--4132, Feb. 2017.

\bibitem{2010ASSP...14..127R}
J.~A. {Rubi{\~n}o-Mart{\'{\i}}n}, R.~{Rebolo}, M.~{Tucci},
  R.~{G{\'e}nova-Santos}, S.~R. {Hildebrandt}, R.~{Hoyland}, J.~M. {Herreros},
  F.~{G{\'o}mez-Re{\~n}asco}, C.~L. {Caraballo},
  E.~{Mart{\'{\i}}nez-Gonz{\'a}lez}, P.~{Vielva}, D.~{Herranz}, F.~J. {Casas},
  E.~{Artal}, B.~{Aja}, L.~d. {Fuente}, J.~L. {Cano}, E.~{Villa},
  A.~{Mediavilla}, J.~P. {Pascual}, L.~{Piccirillo}, B.~{Maffei}, G.~{Pisano},
  R.~A. {Watson}, R.~{Davis}, R.~{Davies}, R.~{Battye}, R.~{Saunders},
  K.~{Grainge}, P.~{Scott}, M.~{Hobson}, A.~{Lasenby}, G.~{Murga},
  C.~{G{\'o}mez}, A.~{G{\'o}mez}, J.~{Ari{\~n}o}, R.~{Sanquirce}, J.~{Pan},
  A.~{Vizcarg{\"u}enaga}, and B.~{Etxeita}, ``{The QUIJOTE CMB Experiment},''
  \emph{Astrophysics and Space Science Proceedings}, vol.~14, p. 127, 2010.

\bibitem{2012SPIE.8444E..2YR}
J.~A. {Rubi{\~n}o-Mart{\'{\i}}n}, R.~{Rebolo}, M.~{Aguiar},
  R.~{G{\'e}nova-Santos}, F.~{G{\'o}mez-Re{\~n}asco}, J.~M. {Herreros}, R.~J.
  {Hoyland}, C.~{L{\'o}pez-Caraballo}, A.~E. {Pelaez Santos}, V.~{Sanchez de la
  Rosa}, A.~{Vega-Moreno}, T.~{Viera-Curbelo}, E.~{Mart{\'{\i}}nez-Gonzalez},
  R.~B. {Barreiro}, F.~J. {Casas}, J.~M. {Diego}, R.~{Fern{\'a}ndez-Cobos},
  D.~{Herranz}, M.~{L{\'o}pez-Caniego}, D.~{Ortiz}, P.~{Vielva}, E.~{Artal},
  B.~{Aja}, J.~{Cagigas}, J.~L. {Cano}, L.~{de la Fuente}, A.~{Mediavilla},
  J.~V. {Ter{\'a}n}, E.~{Villa}, L.~{Piccirillo}, R.~{Battye}, E.~{Blackhurst},
  M.~{Brown}, R.~D. {Davies}, R.~J. {Davis}, C.~{Dickinson}, S.~{Harper},
  B.~{Maffei}, M.~{McCulloch}, S.~{Melhuish}, G.~{Pisano}, R.~A. {Watson},
  M.~{Hobson}, K.~{Grainge}, A.~{Lasenby}, R.~{Saunders}, and P.~{Scott},
  ``{The QUIJOTE-CMB experiment: studying the polarisation of the galactic and
  cosmological microwave emissions},'' in \emph{Ground-based and Airborne
  Telescopes IV}, ser. Proceedings of the SPIE, vol. 8444, Sep. 2012, p.
  84442Y.

\bibitem{LFI}
N.~Aghanim, C.~Armitage-Caplan, M.~Arnaud, M.~Ashdown, F.~Atrio-Barandela,
  J.~Aumont, C.~Baccigalupi, A.~J.~Banday, R.~Barreiro, E.~Battaner,
  K.~Benabed, A.~Benoît, A.~Benoit-Lévy, J.~P.~Bernard, M.~Bersanelli,
  P.~Bielewicz, J.~Bobin, J.~J.~Bock, A.~Bonaldi, and A.~Zonca, ``Planck 2013
  results. iii. lfi systematic uncertainties,'' \emph{Astronomy and
  Astrophysics}, vol. 571, 03 2013.

\bibitem{2017ITAP...65..644K}
K.~{Kundert}, U.~{Rau}, E.~{Bergin}, and S.~{Bhatnagar}, ``{Understanding
  Systematic Errors Through Modeling of ALMA Primary Beams},'' \emph{IEEE
  Transactions on Antennas and Propagation}, vol.~65, pp. 644--653, Feb. 2017.

\bibitem{2012RaSc...47.0K06R}
A.~E.~E. {Rogers} and J.~D. {Bowman}, ``{Absolute calibration of a wideband
  antenna and spectrometer for accurate sky noise temperature measurements},''
  \emph{Radio Science}, vol.~47, p. RS0K06, Aug. 2012.

\bibitem{Winkel}
B.~Winkel, A.~Kraus, and U.~Bach, ``Unbiased flux calibration for single-dish
  telescopes using advanced observing techniques,'' 05 2015, pp. 1--1.

\bibitem{Miguel_Hern_ndez_2019}
\BIBentryALTinterwordspacing
J.~De\;Miguel-Hern{\'{a}}ndez and R.~Hoyland, ``Fundamentals of horn antennas
  with low cross-polarization levels for radioastronomy and satellite
  communications,'' \emph{Journal of Instrumentation}, vol.~14, no.~08, pp.
  R08\,001--R08\,001, aug 2019. [Online]. Available:
  \url{https://doi.org/10.1088%2F1748-0221%2F14%2F08%2Fr08001}
\BIBentrySTDinterwordspacing

\bibitem{EAST}
P.~W. East, ``Fifty years of instantaneous frequency measurement,'' \emph{IET
  Radar, Sonar \& Navigation}, vol.~6, pp. 112--122, Feb. 2012.

\bibitem{RHODES}
J.~D. {Rhodes}, ``{New technique improves IFM performance},'' \emph{Microwaves
  \& RF}, vol.~24, pp. 121--125, Mar. 1985.

\bibitem{IFM}
M.~J. Thornton, ``Ultra-broadband frequency discriminator designs for ifm
  receivers,'' \emph{IEE Colloquium on Multi-Octave Active and Passive
  Components and Antennas}, pp. 13/1 -- 13/4, May 1989.

\bibitem{THORNTON}
M.~Thornton, ``Frequency discriminators for broadband applications,'' in
  \emph{Proc. {Automated RF \& Microwave Measurement Society}}, 2011.

\bibitem{fellgett1951theory}
P.~B. Fellgett, \emph{{Theory of Infra-Red Sensitivities and its Application to
  Investigations of Stellar Radiation in the Near Infra-Red}}.\hskip 1em plus
  0.5em minus 0.4em\relax Reading, U.K.: University of Reading, 1949.

\bibitem{2012SPIE.8452E..33H}
R.~J. {Hoyland}, M.~{Aguiar-Gonz{\'a}lez}, B.~{Aja}, J.~{Ari{\~n}o},
  E.~{Artal}, R.~B. {Barreiro}, E.~J. {Blackhurst}, J.~{Cagigas}, J.~L. {Cano
  de Diego}, F.~J. {Casas}, R.~J. {Davis}, C.~{Dickinson}, B.~E. {Arriaga},
  R.~{Fernandez-Cobos}, L.~{de la Fuente}, R.~{G{\'e}nova-Santos},
  A.~{G{\'o}mez}, C.~{Gomez}, F.~{G{\'o}mez-Re{\~n}asco}, K.~{Grainge},
  S.~{Harper}, D.~{Herran}, J.~M. {Herreros}, G.~A. {Herrera}, M.~P. {Hobson},
  A.~N. {Lasenby}, M.~{Lopez-Caniego}, C.~{L{\'o}pez-Caraballo}, B.~{Maffei},
  E.~{Martinez-Gonzalez}, M.~{McCulloch}, S.~{Melhuish}, A.~{Mediavilla},
  G.~{Murga}, D.~{Ortiz}, L.~{Piccirillo}, G.~{Pisano}, R.~{Rebolo-L{\'o}pez},
  J.~A. {Rubi{\~n}o-Martin}, J.~L. {Ruiz}, V.~{Sanchez de la Rosa},
  R.~{Sanquirce}, A.~{Vega-Moreno}, P.~{Vielva}, T.~{Viera-Curbelo},
  E.~{Villa}, A.~{Vizcarg{\"u}enaga}, and R.~A. {Watson}, ``{The status of the
  QUIJOTE multi-frequency instrument},'' in \emph{Millimeter, Submillimeter,
  and Far-Infrared Detectors and Instrumentation for Astronomy VI}, ser.
  Proceedings of the SPIE, vol. 8452, Sep. 2012, p. 845233.

\bibitem{7067876}
B.~G.~M. de~Oliveira, M.~T. de~Melo, I.~Llamas-Garro, M.~Espinosa-Espinosa,
  M.~R.~T. de~Oliveira, and E.~M.~F. de~Oliveira, ``Integrated instantaneous
  frequency measurement subsystem based on multi-band-stop filters,'' in
  \emph{2014 Asia-Pacific Microwave Conference}, Nov 2014, pp. 910--912.

\bibitem{Badran}
H.~Badran, ``A low cost instantaneous frequency measurement system,''
  \emph{Progress In Electromagnetics Research M}, vol.~59, 08 2017.

\bibitem{8467497}
H.~Rahimpour and N.~Masoumi, ``Design and implementation of a high-sensitivity
  and compact-size ifm receiver,'' \emph{IEEE Transactions on Instrumentation
  and Measurement}, pp. 1--8, 2018.

\bibitem{Rahimpour}
R.~Hamid and N.~Masoumi, ``High-resolution frequency discriminator for
  instantaneous frequency measurement subsystem,'' \emph{IEEE Transactions on
  Instrumentation and Measurement}, vol.~PP, pp. 1--9, 04 2018.

\bibitem{souza}
M.~F.~A.~de Souza, F.~R.~L.~e Silva, T.~M. T.~d. Melo, and L.~Novo,
  ``Discriminators for instantaneous frequency measurement subsystem based on
  open-loop resonators,'' \emph{Microwave Theory and Techniques, IEEE
  Transactions on}, vol.~57, pp. 2224 -- 2231, 10 2009.

\bibitem{TOOLS}
T.~L. Wilson, K.~Rohlfs, and S.~Hüttemeister, \emph{Tools of Radio Astronomy},
  6th~ed.\hskip 1em plus 0.5em minus 0.4em\relax Berlin, Germany:
  Springer-Verlag Berlin Heidelberg, 2013.

\bibitem{KIM}
K.~Nam-Tae, ``Ultra-wideband bias-tee design using distributed network
  synthesis,'' \emph{IEICE Electronics Express}, vol.~10, no.~15, pp.
  20\,130\,472--20\,130\,472, Feb. 2013.

\bibitem{1964pecm.book.....G}
P.~E. {Gray}, D.~{Dewitt}, and A.~R. {Boothroyd}, \emph{{Physical electronics
  and circuit models of transistors}}, 1964.

\bibitem{Master}
J.~De~Miguel~Hern\'andez, \emph{Design of Components for the New Microwave
  Frequency Spectrometer of the Quijote Experiment}.\hskip 1em plus 0.5em minus
  0.4em\relax Universidad de La Laguna, 2017.

\end{thebibliography}

\end{document}